\documentclass[preprint,showpacs,preprintnumbers,amsmath,amssymb]{revtex4}
\usepackage{epsfig}
\usepackage{graphicx}
\usepackage{dcolumn}
\usepackage{bm}
\def\lsim{\mathrel {\vcenter {\baselineskip 0pt \kern 0pt
    \hbox{$&lt;$} \kern 0pt \hbox{$\sim$} }}}
\def\gsim{\mathrel {\vcenter {\baselineskip 0pt \kern 0pt
    \hbox{$&gt;$} \kern 0pt \hbox{$\sim$} }}}

\newcommand{\U}{{\cal {U}}}

\newcommand{\YD}{\rm {YD}}

\newcommand{\x}{\times10^\text}

\begin{document}

\title{Constraints on Unparticle Interaction from $b\to s \gamma$}

\author{Xiao-Gang He and Lu-Hsing Tsai}
\affiliation{Department of Physics
and Center for Theoretical Sciences, National Taiwan University,
Taipei, Taiwan, R.O.C}

\date{\today}

\begin{abstract}
We study unparticle effects on $b\to s \gamma$. The unparticle contributions can contribute
significantly to both left- and right-handed chirality amplitudes. Using available experimental data  and SM calculation for
$B\to X_s \gamma$,
we obtain constraints on various vector and scalar unparticle couplings.
We find that the constraints sensitively depend on the unparticle dimension $d_\U$. For $d_\U$ close
to one, the constraints can be very stringent. The constraints become weak when $d_\U$
is increased. In general the constraints on scalar unparticle couplings are
weaker than those for vector unparticle couplings. Sizeable coupling strength for unparticles with quarks
is still allowed. We also show that polarization measurement in $\Lambda_b \to \Lambda \gamma$ can further
constrain the couplings.
\end{abstract}


\maketitle
\section{Introduction}

Recently Georgi proposed an interesting idea to describe possible
scale invariant effect at low energies by
unparticles\cite{Georgi:2007ek}. It was argued that operators
$O_{SI}$ made of fields in the scale invariant sector may
interact with operators $O_{SM}$ of dimension $d_{SM}$ made of
Standard Model (SM) fields at some high energy scale by exchanging
particles with large masses, $M_{\cal{U}}$, with the generic form
$O_{SM} O_{SI}/M^k_{\cal{U}}$. At another scale
$\Lambda_{\cal{U}}$ the scale invariant sector induce dimensional
transmutation, below that scale the operator $O_{SI}$ matches
onto an unparticle operator $O_{\cal{U}}$ with dimension $d_\U$ and
the unparticle interaction with SM particles at low energy has the
form \begin{eqnarray} \lambda \Lambda_{\cal{U}} ^{4-d_{SM} - d_\U}
O_{SM} O_{\cal{U}}.\label{eff}
\end{eqnarray}

Study of unparticle effects has drawn a lot of
attentions from more
theoretically related work to more phenomenologically studies. There
are many possible ways unparticles may interact with the SM
particles\cite{Chen:2007qr}. Most of the phenomenological work concentrate on possible
effects of unparticle interactions with SM particles and constraints
on the interaction strength $\lambda/\Lambda_{\U}^{4-d_{SM}-d_\U}$. One of the
subjects where a lot of activities have been devoted to is the study of low energy rare flavor changing processes
involving quarks\cite{quark} and charged leptons\cite{lepton,Ding:2008zza}.
In this work we study unparticle effects on $b\to s \gamma$
and constrain unparticle interactions using known SM values and current experimental data for
$B \to X_s \gamma$.

The rare $b\to s\gamma$ decay process has been shown to provide interesting constraints
on possible new physics beyond the SM\cite{new}. Experimentally the leading contribution
to $B \to X_s \gamma$ with large $\gamma$ energy $E_\gamma$ is dominated by $b\to s\gamma$.
Experimental measurement on this decay has achieved
very high precision with $B(B \to X_s \gamma)$ given by\cite{HFAG:2008}
$(3.52\pm 0.23\pm0.09)\times10^{-4}$,
with $E_\gamma >1.6 \text{GeV}$. On the theoretical side,
the SM calculation for $B(B \to X_s \gamma)$ has been evaluated at the NNLO order
\cite{Misiak:2007} with $(3.15\pm 0.23)\times10^{-4}$ for $E_\gamma > 1.6 \text{GeV}$.
It is clear that experimental data and
SM prediction agree with each other very well leaving small room for new physics beyond the SM.
Taking this on the positive side, the process $B\to X_s \gamma$ can provide stringent constraints
on possible new physics beyond the SM.  Several flavor changing processes have been studied\cite{quark,lepton,Ding:2008zza}, but
unparticle contribution to $b\to s\gamma$ has not been
studied. We therefore concentrate on this subject.

Although at present the detailed dynamics for interaction between unparticles and SM particles are not known,
unparticle effects on various physical processes can be studied from effective theory point of view
using eq.(\ref{eff}). The main task is then, as many phenomenological studies of unparticle physics, to
use available data to constrain the allowed parameter space and to see how large the
unparticle effects can be and to test possible effects experimentally.

In principle, when the unparticle sector is coupled to the SM sector the
scale invariance is broken due to finite mass of the SM fields\cite{Deshpande:2007jy} and also due to spontaneous symmetry breaking
of Higgs vacuum expectation value if coupled\cite{Fox:2007sy}.  The unparticle behavor may only exist in a window below the scale
$\Lambda_\U$ and above a scale $\mu$ where the scale invariance is broken again by SM particle effects.  If this is the case, the contributions
of the unparticles should only be within this window and below $\mu$ the effects should be replace by that resulted from
the residual degrees of freedom. However, at this stage there is no specific way, as far as we know, to describe such effects. In our
study of unparticle effects on $b\to s \gamma$, we will follow most of the phenomenological studies in the literature assuming that the unparticle effects from the scale $\Lambda_\U$ down to zero.

We will study $b\to s\gamma$ using the lowest possible dimension operators
due to scalar and vector unparticle and SM fields interactions. The  unparticle contributions can contribute
significantly to both left- and right-handed chirality amplitudes. Using available experimental data and SM calculation mentioned above,
we obtain constraints on various vector and scalar unparticle couplings.
We find that the constraints sensitively depend on the unparticle dimension $d_\U$. For $d_\U$ close
to one, the constraints can be very stringent. The constraints become weak when $d_\U$
is increased. In general the constraints on scalar unparticle couplings are
weaker than those for vector unparticle couplings. Sizeable coupling strength for unparticles with quarks
is still allowed. We also find that polarization measurement in $\Lambda_b \to \Lambda \gamma$ can further
constraint the couplings.

\section{Unparticle contribution to $b\to s\gamma$}

The lowest dimension operators, which can generate contributions to $b \to s \gamma$ at one loop level,
come from interaction of vector unparticle with quarks and are given by\cite{Chen:2007qr}
\begin{eqnarray*}
\lambda'_{QQ}\Lambda_{\cal{U}}^{1-d_\U}\bar Q_L \gamma_\mu Q_L O^\mu_{\cal{U}},\;\;
\;\;\lambda'_{DD}\Lambda_{\cal{U}}^{1-d_\U}\bar D_R  \gamma_\mu D_R O^\mu_{\cal{U}}.
\end{eqnarray*}
Here $Q_L = (U_L, D_L)^T$, $D_R$ are the SM left-handed quark
doublet, and right-handed down-quark, respectively.

Scalar unparticle interaction with quarks can also induce $b  \to s \gamma$.
The lowest dimension operators which can contribute to $b\to s\gamma$ is at order $\Lambda_\U^{-d_{\cal{U}}}$.
The following operator will generate finite
contributions to $b\to s\gamma$ at one loop level\cite{Chen:2007qr}
\begin{eqnarray}
\lambda^{\rm{YD}}\Lambda_{\cal{U}}^{-d_\U}\bar Q_L \tilde H D_R O_{\cal{U}},
\end{eqnarray}
where $H$ is the SM Higgs doublet transforming under the SM gauge group $SU(2)_L\times U(1)_Y$ as $(2, 1)$.

After the Higgs develops a non-zero vacuum expectation value $<H> = v$,
the above interaction between quarks and an unparticle becomes
\begin{eqnarray}
\lambda^{\rm{YD}}_{sb}v\Lambda_{\cal{U}}^{-d_\U}\bar s_L b_R
O_{\cal{U}}.
\end{eqnarray}

At the same order in $\Lambda_\U$, there are several other operators involving
quarks and a scalar unparticle, such as\cite{Chen:2007qr} $\bar Q_{L} \gamma_\mu
\partial^\mu Q_{L} O_{\cal{U}}$,  $\bar D_{R} \gamma_\mu
\partial^\mu D_{R} O_{\cal{U}}$, $\bar Q_{L}  \gamma_\mu Q_{L}
\partial^\mu O_{\cal{U}}$, and $\bar D_{R}  \gamma_\mu D_{R}
\partial^\mu O_{\cal{U}}$. However, their one loop contributions to $b \to s \gamma$ diverge due to
derivative couplings. Additional parameters or operators are need to render these divergences making the
effects not calculable. We will not consider their effects here.

\begin{figure}
\graphicspath{{images/}}
\includegraphics[width=9cm]{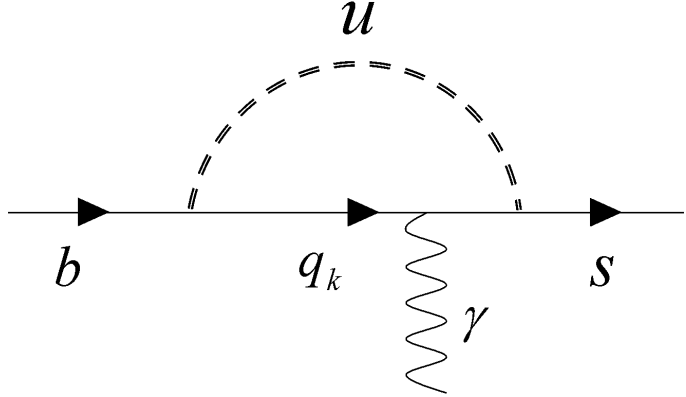}
\caption{One loop Feynmann diagram for $b \to s \gamma$ by exchanging an unparticle.
The $q_k$ can be $d$, $s$, and $b$ quark }\label{b2sgamma}
\end{figure}

The one loop Feynmann diagram  giving contribution to $b\to s \gamma$ is shown in Fig. 1.
We will indicate the incoming $b$ quark by $q_i$ and the out going $s$ quark by $q_j$.
The formula obtained can be easily adapted for other incoming and out going fermions.
We obtain the vector unparticle contribution to
$q_i\to q_j \gamma$ amplitude as
\begin{eqnarray}
\mathcal{M}_v(q_i\to q_j \gamma)
=\frac{Qe}{2m_i}N_v(d_\U)\bar{q_{j}}i\sigma_{\mu\nu}\epsilon^{*\mu}q^{\nu}(A^L_v
L+A^R_v R)q_i, \label{vector}
\end{eqnarray}
where $q$ and $\epsilon^\mu$ are the photon momentum and polarization, respectively.
$N_v(d_\U)= (A_{d_\U}/{16\pi^2\sin(\pi
d_\U))(m_i/\Lambda_u)^{2d_\U-2}}$ with $A_{d_\U} = (16\pi^{5/2}/(2\pi)^{2d_\U})
\Gamma(d_\U + 1/2)/\Gamma(d_\U -1) \Gamma(2 d_\U)$,
and
\begin{eqnarray}
A^R_v&=&[-2f_1+2f_2-f_{2x}+\frac{1}{2-d_\U}(f_1-3f_{2x})+z_{ki}^2(f_3-2f_{3x})]
\lambda^L_{jk}\lambda^L_{ki}\nonumber\\
&+&z_{ji}[-2f_1+f_2+f_{2x}+\frac{1}{2-d_\U}(f_1-3f_2+3f_{2x})-z_{ki}^2(f_3-2f_{3x})]
\lambda^R_{jk}\lambda^R_{ki}\nonumber\\
&+&z_{ki}[4f_1-f_2+\frac{1}{2-d_\U}(3f_2-2f_1)+z_{ki}^2f_3]\lambda^L_{jk}\lambda^R_{ki} - f_3z_{ji} z_{ki}
\lambda^R_{jk}\lambda^L_{ki}\nonumber\\
A^L_v&=&[-2f_1+2f_2-f_{2x}+\frac{1}{2-d_\U}(f_1-3f_{2x})+z_{ki}^2(f_3-2f_{3x})]
\lambda^R_{jk}\lambda^R_{ki}\nonumber\\
&+&z_{ji}[-2f_1+f_2+f_{2x}+\frac{1}{2-d_\U}(f_1-3f_2+3f_{2x})-z_{ki}^2(f_3-2f_{3x})]
\lambda^L_{jk}\lambda^L_{ki}\nonumber\\
&+&z_{ki}[4f_1-f_2+\frac{1}{2-d_\U}(3f_2-2f_1)+z_{ki}^2f_3]\lambda^R_{jk}\lambda^L_{ki}
- f_3z_{ji} z_{ki}\lambda^L_{jk}\lambda^R_{ki},
\end{eqnarray}
where $z_{ji}= m_j/m_i$.  $\lambda^L = \lambda'_{QQ}$,
and $\lambda^R = \lambda'_{DD}$. The functions $f_i(d_\U)$ are defined as
\begin{eqnarray}
f_0(d_\U)&=&\int_0^{1}dx\int_0^{1}dy
\frac{y^{d_\U}(1-y)^{1-d_\U}}{u^{2-d_\U}}, \;\;
f_1(d_\U)= \int_0^{1}dx\int_0^{1}dy
\frac{y^{d_\U-1}(1-y)^{2-d_\U}}{u^{2-d_\U}}\nonumber\\
f_2(d_\U)&=&\int_0^{1}dx\int_0^{1}dy
\frac{y^{d_\U}(1-y)^{2-d_\U}}{u^{2-du}},\;\;
f_{2x}(d_\U)= \int_0^{1}dx\int_0^{1}dy
\frac{xy^{d_\U}(1-y)^{2-d_\U}}{u^{2-du}}\nonumber\\
f_3(d_\U)&=&\int_0^{1}dx\int_0^{1}dy
\frac{y^{d_\U}(1-y)^{2-d_\U}}{u^{3-d_\U}},\;\;
f_{3x}(d_\U)=\int_0^{1}dx\int_0^{1}dy
\frac{xy^{d_\U}(1-y)^{2-d_\U}}{u^{3-d_\U}},
\end{eqnarray}\\
with $u = z_{ki}^2-(1-x)(1-y)-x(1-y)z_{ji}^2$.

For scalar contribution, we obtain
\begin{eqnarray}
\mathcal{M}_s(q_i \to q_j \gamma)
=\frac{Qe}{2m_i}N_s(d_\U)\bar{q_{j}}i\sigma_{\mu\nu}\epsilon^{*\mu}q^{\nu}(A^L_s
L+A^R_s R)q_i, \label{scalar}
\end{eqnarray}
where $N_s(d_\U)= (A_{d_\U}/16\pi^2\sin(\pi
d_\U))(v^2/m_i^2)(m_i/\Lambda_u)^{2d_\U}$, and
\begin{eqnarray}
A^R_s&=&[f_2(d_\U)-f_{2x}(d_\U)]\lambda^{\YD}_{jk}\lambda^{\YD *}_{ik}
+f_{2x}(d_\U)z_{ji}\lambda^{\YD*}_{kj}\lambda^{\YD}_{ki}+f_0(d_\U)z_{ki}\lambda^{\YD}_{jk}\lambda^{\YD}_{ki}\\
A^L_s&=&[f_2(d_\U)-f_{2x}(d_\U)]\lambda^{\YD*}_{kj}\lambda^{\YD}_{ki}
+f_{2x}(d_\U)z_{ji}\lambda^{\YD}_{jk}\lambda^{\YD*}_{ik}+f_0(d_\U)z_{ki}\lambda^{\YD*}_{kj}\lambda^{\YD*}_{ik}.
\end{eqnarray}
In our calculations, we will use the central quark masses given in PDG\cite{pdg}: $m_b = 4.70$ GeV, $m_s = 95$ MeV and $m_s/m_d = 19$.

The above amplitudes are evaluated at the unparticle scale $\mu = \Lambda_\U$. When running down to the relevant scale
$\mu = m_b$ for $b\to s \gamma$, there are corrections. The gluonic penguin with the photon $\gamma$
replaced by  a gluon $g$, $b\to s g$, generated at $\Lambda_\U$ will also contribute to $b\to s\gamma$ at a lower scale $m_b$. The amplitude for $M_{\U = v,s}(q_i \to q_j g)$ is given by
\begin{eqnarray}
&&\mathcal{M}_\U (q_i\to q_j g)
=\frac{g_s}{2m_i}N_v(d_\U)\bar{q_{j}}i\sigma_{\mu\nu}\epsilon_a^{*\mu}q^{\nu}(A^L_\U
L+A^R_\U R)T^aq_i,
\end{eqnarray}
where $g_s$ is the strong interaction coupling, $\epsilon^\mu_a$ is the gluon polarization vector
and $T^a$ is the generator of the
color gauge group $SU(3)_C$ normalized to $Tr(T^aT^b) = \delta^{ab}/2$.

One can easily translate the above amplitudes into the usual amplitudes defined by
\begin{eqnarray}
&&M(b\to s \gamma) =-V_{tb}V^*_{ts}{G_F\over \sqrt{2}} {e\over 8 \pi^2} C_7(\mu) \bar s \sigma_{\mu\nu}F^{\mu\nu} (m_s L + m_bR)b,  \nonumber\\
&&M(b\to s g) = -V_{tb}V^*_{ts}{G_F\over \sqrt{2}}{g_s\over 8 \pi^2} C_8(\mu) \bar s \sigma_{\mu\nu} G^{\mu\nu}_a (m_s L + m_b R)T^a b,
\end{eqnarray}
where $F^{\mu\nu}$ and $G^{\mu\nu}_a$ are the field strength of photon and gluon fields.

Using the leading QCD corrected effective Wilson coefficient at the scale $m_b$ for $b\to s\gamma$ is given by\cite{Buras:1993xp}, $C^{eff}_7(m_b) = 0.689C_7(m_W) +0.087C_8(m_W)$,
we obtain an approximate expression for the QCD corrected unparticle contribution, at the scale $\mu = m_b$,
\begin{eqnarray}
&&\mathcal{\tilde M}_\U(q_i \to q_j \gamma)
=\bar{q_{j}}i\sigma_{\mu\nu}\epsilon^{*\mu}q^{\nu}(\tilde A^L_\U
L+\tilde A^R_\U R)q_i, \nonumber\\
&&\tilde A^{L,R}_{\U = v,s} = \frac{Qe}{2m_i}N_{\U=v,s}(d_\U)(0.689 + 0.087/Q) A^{L,R}_{\U = v,s}.
\end{eqnarray}

Using the above expression one can put constraints on unparticle couplings.

\section{Numerical Analysis and Conclusions}

To see how unparticle interactions affect $B\to X_s \gamma$, we
use the following to measure possible unparticle contribution,
\begin{eqnarray}
R_{exp-SM}=\frac{\Gamma_{exp}-\Gamma_{SM}}{\Gamma_{SM}}= \frac{B_{exp}}{B_{SM}}-1.
\end{eqnarray}

Using the available experimental and SM
values, we find $R_{exp-SM} = 0.117 \pm 0.113$ with $E_\gamma > 1.6$ GeV. It is clear that
at this stage there  is no evidence of new physics beyond SM. However,
we can turn the argument around and use allowed value of $R$ to constrain new interactions.

To compare with data and aim at the leading correction from
unparticle to the SM prediction, we first define an effective SM for $b\to s\gamma$ amplitude $\tilde A_{SM}^{L,R}$
with $\tilde A_{SM}^L/\tilde A_{SM}^R \approx m_s/m_b$, as should be in the SM,  such that the corresponding
Wilson coefficient at the leading order amplitude\cite{Buras:1993xp} reproduces the SM prediction for
the branching ratio with relevant in put parameters from Ref.\cite{pdg}. We
then add to it the leading QCD corrected unparticle contribution $\tilde A^{L,R}_{\U}$
to obtain the total amplitude. Replacing $\Gamma_{exp}$ by $\Gamma_{un-SM}$ determined by
the total SM and unparticle leading contributions, we obtain a quantity similar to $R_{exp-SM}$
\begin{eqnarray}
R_{un-SM} =
\frac{{\vert \tilde A_{SM}^L+\tilde A_{\U}^L \vert}^2 +\vert {\tilde A_{SM}^R+\tilde A_{\U}^R \vert}^2}{{\vert \tilde A_{SM}^L \vert}^2
+\vert {\tilde A_{SM}^R \vert}^2}-1.
\end{eqnarray}
We finally approximate $R_{un-SM}$ to
$R_{exp-SM}$ and obtain constraints on unparticle couplings.
There are higher order SM corrections to
the above formula, but for our purpose of obtaining leading
constraints on unparticle effects, this should be sufficient.

In the SM $\tilde A^L_{SM}/\tilde A^R_{SM} = m_s/m_b$. It is obvious that the
main contribution of SM is the right hand couplings. For
unparticle contributions, $\tilde A^L_{\U}$ can be comparable or even
larger than $\tilde A^R_{\U}$. We will obtain bounds on the unparticle conuplings from
data and known SM numbers allow the theoretical value $R_{un-SM}$ to be in
the $1\sigma$ range. Depending on the intermediate quarks
exchanged in the loop, different quark-unparticle couplings can
appear. We will constrain the coupling for each of the
combinations with non-zero contribution and set other
equal to zero first assuming the couplings are all real.

There are three possibilities involving a quark in the loop. We discuss them in the following.

a) For $d$ quark and vector unparticle in the loop, it is possible to constrain
$\lambda^R_{s d}\lambda^R_{d b}$, $\lambda^L_{s d}\lambda^L_{d b}$, $\lambda^R_{s d}\lambda^L_{d b}$,
$\lambda^L_{s d}\lambda^R_{d b}$.
For scalar unparticle in the loop, it is possible to constrain $\lambda^{\YD}_{s d}\lambda^{\YD*}_{b d}$,
$\lambda^{\YD*}_{ds}\lambda^{\YD}_{db}$,
$\lambda^{\YD}_{s d}\lambda^{\YD}_{db}$, and $\lambda^{\YD*}_{ds}\lambda^{\YD*}_{b d}$.

b)  For $s$ quark and vector unparticle in the loop, it is possible to constrain
$\lambda^R_{s s}\lambda^R_{s b}$, $\lambda^L_{s s}\lambda^L_{s b}$, $\lambda^R_{s s}\lambda^L_{s b}$
and $\lambda^L_{s s}\lambda^R_{s b}$.
For scalar unparticle in the loop, it is possible to constrain $\lambda^{\YD}_{s s}\lambda^{\YD*}_{b s}$,
$\lambda^{\YD}_{s s}\lambda^{\YD}_{sb}$, $\lambda^{\YD*}_{s s}\lambda^{\YD}_{sb}$,
and $\lambda^{\YD*}_{s s}\lambda^{\YD*}_{bs }$. For real couplings, there are only two needed to be
considered, with sub-indices $(ss,sb)$ and $(ss,bs)$.

c)  For $b$ quark and vector unparticle in the loop, it is possible to constrain
$\lambda^R_{s b}\lambda^R_{b b}$, $\lambda^L_{s b}\lambda^L_{b b}$, $\lambda^R_{s b}\lambda^L_{b b}$
and $\lambda^L_{s b}\lambda^R_{b b}$.
For scalar unparticle in the loop, it is possible to constrain $\lambda^{\YD}_{s b}\lambda^{\YD*}_{b b}$,
$\lambda^{\YD}_{s b}\lambda^{\YD}_{bb}$, $\lambda^{\YD*}_{b s}\lambda^{\YD}_{bb}$,
and $\lambda^{\YD*}_{b s}\lambda^{\YD*}_{bb }$. For real couplings, there are only two needed to be considered,
with sub-indices $(sb,bb)$ and $(bs,bb)$.

The various constraints are shown in Tables I, II and III for different values of unparticle dimension
$d_\U$ with
$\Lambda_\U$ set to be 1 TeV. The central values
for the unparticle couplings are obtained by taking the SM leading
values and require the unparticle contributions to produce the
central value of $R_{exp-SM}$. In general there are two solutions. One
comes from constructive interference contribution relative to the SM
dominant contribution, and another
from destructive interference. In the case that
the unparticle contribution is dominated by the same chirality,
$R = (1+\gamma_5)/2$, amplitude as that of the
dominate one in SM,
the allowed unparticle amplitude from destructive case will be larger than the SM
one.
These are the cases with one of the central values
(absolute values) much larger than the other in the tables. We hold
the view that SM should dominate the contribution
to $b\to s\gamma$,
therefore we consider these cases not good ones for constraints.

For bounds on the couplings, we list the bounds corresponding to positive and negative solutions
separately in the same way as their central values.
Positive numbers indicate that the couplings should be
smaller than the numbers listed, and negative
numbers indicate that the couplings should be larger than the numbers listed.

It can be seen that the constraints sensitively depend on the unparticle dimension
parameter $d_\U$. For $d_\U$ not too far away from 1, the constraints
are stringent, but become weaker as $d_\U$ increases. It is also
clear that the constraints on the vector unparticle couplings are
stronger than those for scalar unparticle couplings. This can be
easily understood by noticing that the scalar unparticle couplings
is suppressed by a factor of $v/\Lambda_\U$ compared with vector unparticle
couplings. Sizeable coupling strength for unparticles with quarks
is still allowed.

Note that using $B\to X_s\gamma$ branching ratio alone, it is not possible to
distinguish the above solutions since it is proportional to $|\tilde A_R^{\mbox{total}}|^2 + |\tilde A_L^{\mbox{total}}|^2$
which is how the constraints are obtained.
We comment that measurement of polarization $\alpha_\Lambda$ in $\Lambda_b \to \Lambda \gamma$ can provide
more information to distinguish some of the solutions. The polarization parameter $\alpha_\Lambda$ is defined by\cite{polarization}
\begin{eqnarray}
&&{d\Gamma \over \Gamma d \cos\theta} = {1\over 2} (1+\alpha_\Lambda \cos\theta)\;,\;\;\;\;\alpha_\Lambda ={|\tilde A_R^{\mbox{total}}|^2-|\tilde A_L^{\mbox{total}}|^2\over |\tilde A_R^{\mbox{total}}|^2 + |\tilde A_L^{\mbox{total}}|^2}\;,
\end{eqnarray}
where $\Gamma$ is the decay rate for $\Lambda_b \to \Lambda \gamma$, and $\theta$
is the angle between the $\Lambda$ polarization and the photon momentum directions.

In the SM, since $\tilde A_{SM}^L/\tilde A^R_{SM} = m_s/m_b$, one would have $\alpha_\Lambda \approx 1$.
In the Tables, we list $\alpha_\Lambda$ for the corresponding constraints on the couplings. We see that unparticle contributions can change
the value for $\alpha_\Lambda$ significantly. Future measurement for $\alpha_\Lambda$ can provide more information about unparticle interactions.

There are several studies of unparticle flavor changing effects in $B$ decays. The couplings
are constrained from several processes\cite{quark}, such as
stringent constraints on the couplings
$(\lambda'_{(d,s)b})^2$ and $(\lambda^{YD}_{(d,s)b} - \lambda^{YD*}_{b(d,s)})^2$
from $B_{d,s}-\bar B_{d,s}$ mixing\cite{quark}. If all $\lambda'_{ij}$ (or $\lambda^{YD}_{ij}$) are similar in size, the constraints
from these considerations are stronger than the ones obtained on Tables I, II and III. However, one cannot exclude that the couplings
are different for different generations, therefore the constraints obtained here are on different combinations and are new.
There are also several studies of radiative decays involving
leptons\cite{Ding:2008zza}. The couplings obtained here are in general
less stringent compared with the ones involving leptons. The bounds obtained here involve quarks and are, again, new ones.

In conclusion, we have studied unparticle effects on
$b\to s \gamma$. The unparticle contributions can contribute significantly to
both left- and right-handed chirality amplitudes. Using available experimental data on $b\to s \gamma$
and SM calculation,
we have obtained new constraints on various vector and scalar unparticle couplings.
The constraints sensitively depend on the unparticle dimension $d_\U$. For $d_\U$ close
to one, the constraints can be very stringent. The constraints become weaker when $d_\U$
is increased. In general the constraints on scalar unparticle couplings are
weaker than those for vector unparticles. Sizeable coupling strength for scalar unparticles
are still allowed leaving rooms for direct search for unparticle effects at colliders,
such as LHC. Polarization measurement in $\Lambda_b \to \Lambda \gamma$ can further
constraint the couplings.

\vskip 1.0cm \noindent {\bf Acknowledgments}$\,$ We thank Shao-Long Chen for early participation in
this work and for many discussions. This work was supported in part by the NSC and NCTS.

\newpage

\begin{table}[h]
\caption{Central values (c-value)  and bounds for unparticle couplings with $d$ quark in
the loop for $\Lambda_\U =$ 1 TeV. In the table ``-'' indicates that the central values are larger than
10 implying weak constraints which we do not list. The corresponding values for $\alpha_\Lambda$ are listed
below the constraints on couplings. In the table ``$\sim 1$.'' indicates a value very close to one.}
\vspace{0.5cm}{\scriptsize
$\begin{array}{|c|ccccc|}
\hline
d_\U & 1.1 & 1.3 & 1.5 & 1.7 & 1.9 \\
\hline

 \text{$\lambda^R_{sd}\lambda^R_{db}$ c-value}&-2.1\x{-5}(2.4\x{-5})&-0.00073(0.00079)&-0.019(0.019)&-
0.30(0.28)&-1.3(1.2)\\
\text{$\alpha_\Lambda$}&0.79(0.79)&0.79(0.79)&0.79(0.79)&0.79(0.79)&0.79(0.79)
\\
\text{$\lambda^R_{sd}\lambda^R_{db}$ bound}&-3.0\x{-5}(3.2\x{-5})&-0.0010(0.0011)&-0.026(0.026)&-0.41(
0.39)&-1.8(1.6)\\
\text{$\alpha_\Lambda$}&0.63(0.62)&0.63(0.62)&0.63(0.63)&0.62(0.63)&0.62(0.63)
\\
\hline

\text{$\lambda^L_{sd}\lambda^L_{db}$ c-value}&-3.9\x{-6}(0.00013)&-0.00021(0.0028)&-0.019(0.019)&-1.
1(0.077)&-7.0(0.21)\\
\text{$\alpha_\Lambda$}&\sim 1.(1.)&\sim 1.(1.)&\sim 1.(1.)&\sim 1.(1.)&\sim 1.(1.)\\
\text{$\lambda^L_{sd}\lambda^L_{db}$ bound}&-7.4\x{-6}(0.00013)&-0.00038(0.0030)&-0.026(0.026)&-1.1(0.
14)&-7.2(0.41)\\
\text{$\alpha_\Lambda$}&\sim 1.(1.)&\sim 1.(1.)&\sim 1.(1.)&\sim 1.(1.)&\sim 1.(1.)\\
\hline

\text{$\lambda^R_{sd}\lambda^L_{db}$ c-value}&-0.0056(0.00035)&-0.12(0.068)&-1.6(14)&-&-\\
\text{$\alpha_\Lambda$}&0.98(\sim 1.)&0.99(0.99)&1.(0.84)&-&-\\
\text{$\lambda^R_{sd}\lambda^L_{db}$ bound}&-0.0059(0.00065)&-0.15(0.10)&-2.9(16)&-&-\\
\text{$\alpha_\Lambda$}&0.98(\sim 1.)&0.99(0.99)&0.99(0.83)&-&-
\\
\hline

\text{$\lambda^L_{sd}\lambda^R_{db}$ c-value}&-0.0019(0.0010)&-0.12(0.068)&-4.6(4.9)&-&-\\
\text{$\alpha_\Lambda$}&0.64(0.88)&0.64(0.88)&0.81(0.81)&-&-
\\
\text{$\lambda^L_{sd}\lambda^R_{db}$ bound}&-0.0024(0.0016)&-0.15(0.10)&-6.6(6.8)&-&-\\
\text{$\alpha_\Lambda$}&0.45(0.75)&0.46(0.75)&0.66(0.66)&-&-
\\
\hline

\text{$\lambda^{\YD}_{sd}\lambda^{\YD}_{bd}$ c-value}&-0.048(0.0015)&-0.74(0.054)&-4.2(4.2)&-&-\\
\text{$\alpha_\Lambda$}&0.94(\sim 1.)&0.99(\sim 1.)&\sim 1.(1.)&-&-\\
\text{$\lambda^{\YD}_{sd}\lambda^{\YD}_{bd}$ bound}&-0.049(0.0028)&-0.78(0.099)&-5.9(5.9)&-&-\\
\text{$\alpha_\Lambda$}&0.95(\sim 1.)&0.99(\sim 1.)&\sim 1.(1.)&-&-\\
\hline

\text{$\lambda^{\YD}_{ds}\lambda^{\YD}_{db}$ c-value}&-0.012(0.0061)&-0.23(0.17)&-4.2(4.2)&-&-\\
\text{$\alpha_\Lambda$}&0.63(0.87)&0.74(0.83)&0.79(0.79)&-&-
\\
\text{$\lambda^{\YD}_{ds}\lambda^{\YD}_{db}$ bound}&-0.015(0.0093)&-0.31(0.25)&-5.9(5.9)&-&-\\
\text{$\alpha_\Lambda$}&0.44(0.75)&0.57(0.68)&0.63(0.63)&-&-\\
\hline

\text{$\lambda^{\YD}_{sd}\lambda^{\YD}_{db}$ c-value}&-0.31(0.013)&-4.4(0.79)&-&-&-\\
\text{$\alpha_\Lambda$}&\sim 1.(1.)&\sim 1.(1.)&-&-&-\\
\text{$\lambda^{\YD}_{sd}\lambda^{\YD}_{db}$ bound}&-0.32(0.025)&-5.0(1.4)&-&-&-\\
\text{$\alpha_\Lambda$}&\sim 1.(1.)&\sim 1.(1.)&-&-&-\\
\hline

\text{$\lambda^{\YD}_{ds}\lambda^{\YD}_{bd}$ c-value}&-0.067(0.061)&-1.9(1.8)&-&-&-\\
\text{$\alpha_\Lambda$}&0.79(0.79)&0.79(0.79)&-&-&-
\\
\text{$\lambda^{\YD}_{ds}\lambda^{\YD}_{bd}$ bound}&-0.093(0.087)&-2.7(2.6)&-&-&-\\
\text{$\alpha_\Lambda$}&0.63(0.63)&0.63(0.63)&-&-&-
\\
\hline
\end{array}$}
\label{bds}
\end{table}

\newpage
\begin{table}[h]
 \caption{Central values (c-value)  and bounds for unparticle couplings with $s$ quark in
the loop for $\Lambda_\U =$ 1 TeV. In the table ``-'' indicates that the central values are larger than
10 implying weak constraints which we do not list. The corresponding values for $\alpha_\Lambda$ are listed
below the constraints on couplings. In the table ``$\sim 1$.'' indicates a value very close to one.}
\vspace{0.5cm}{\scriptsize
$\begin{array}{|c|ccccc|}
\hline
d_\U & 1.1 & 1.3 & 1.5 & 1.7 &1.9 \\
\hline

\text{$\lambda^R_{ss}\lambda^R_{sb}$ c-value}&-2.0\x{-5}(2.2\x{-5})&-0.00071(0.00076)&-0.018(0.018)&-
0.29(0.27)&-1.3(1.2)\\
\text{$\alpha_\Lambda$}&0.79(0.79)&0.79(0.79)&0.79(0.79)&0.79(0.79)&0.79(0.79)
\\
\text{$\lambda^R_{ss}\lambda^R_{sb}$ bound}&-2.8\x{-5}(3.1\x{-5})&-0.0010(0.0011)&-0.026(0.026)&-0.41(
0.39)&-1.8(1.6)\\
\text{$\alpha_\Lambda$}&0.63(0.62)&0.63(0.62)&0.63(0.63)&0.62(0.63)&0.62(0.63)
\\
\hline

\text{$\lambda^L_{ss}\lambda^L_{sb}$ c-value}&-4.2\x{-6}(0.00011)&-0.00023(0.0024)&-0.021(0.016)&-1.
1(0.075)&-7.0(0.21)\\
\text{$\alpha_\Lambda$}&\sim 1.(1.)&\sim 1.(1.)&\sim 1.(1.)&\sim 1.(1.)&\sim 1.(1.)\\
\text{$\lambda^L_{ss}\lambda^L_{sb}$ bound}&-7.9\x{-6}(0.00011)&-0.00042(0.0025)&-0.028(0.024)&-1.1(
0.14)&-7.2(0.41)\\
\text{$\alpha_\Lambda$}&\sim 1.(1.)&\sim 1.(1.)&\sim 1.(1.)&\sim 1.(1.)&\sim 1.(1.)\\
\hline

\text{$\lambda^R_{ss}\lambda^L_{sb}$ c-value}&-7.7\x{-5}(0.0019)&-0.0048(0.085)&-0.21(2.5)&-6.2(48)&
-\\
\text{$\alpha_\Lambda$}&\sim 1.(-0.65)&\sim 1.(-1.)&0.99(-0.74)&0.97(-0.3)&-\\
\text{$\lambda^R_{ss}\lambda^L_{sb}$ bound}&-0.00015(0.0020)&-0.009(0.089)&-0.38(2.6)&-&-\\
\text{$\alpha_\Lambda$}&\sim 1.(-0.6)&0.98(-0.99)&0.96(-0.81)&-&-\\
\hline

\text{$\lambda^L_{ss}\lambda^R_{sb}$ c-value}&-0.0011(0.00013)&-0.049(0.0083)&-0.63(0.81)&-4.9(60)&-\\
\text{$\alpha_\Lambda$}&-0.32(0.99)&0.37(0.99)&0.94(0.94)&0.99(0.56)&-\\
\text{$\lambda^L_{ss}\lambda^R_{sb}$ bound}&-0.0012(0.00024)&-0.055(0.014)&-0.92(1.1)&-9.1(64)&-\\
\text{$\alpha_\Lambda$}&-0.43(0.96)&0.28(0.96)&0.89(0.89)&0.98(0.54)&-
\\
\hline

\text{$\lambda^{\YD}_{ss}\lambda^{\YD}_{bs}$ c-value}&-0.021(0.0014)&-0.37(0.050)&-2.8(3.0)&-&-\\
\text{$\alpha_\Lambda$}&0.92(\sim 1.)&\sim 1.(1.)&\sim 1.(1.)&-&-\\
\text{$\lambda^{\YD}_{ss}\lambda^{\YD}_{bs}$ bound}&-0.022(0.0026)&-0.41(0.089)&-4.0(4.2)&-&-\\
\text{$\alpha_\Lambda$}&0.9(\sim 1.)&\sim 1.(1.)&\sim 1.(1.)&-&-\\
\hline

\text{$\lambda^{\YD}_{ss}\lambda^{\YD}_{sb}$ c-value}&-0.015(0.0019)&-0.16(0.11)&-1.3(6.7)&-&-\\
\text{$\alpha_\Lambda$}&0.42(0.98)&0.86(0.92)&0.98(0.54)&-&-\\
\text{$\lambda^{\YD}_{ss}\lambda^{\YD}_{sb}$ bound}&-0.017(0.0033)&-0.22(0.17)&-2.2(7.7)&-&-\\
\text{$\alpha_\Lambda$}&0.33(0.95)&0.76(0.84)&0.94(0.44)&-&-
\\
\hline
\end{array}$}
\label{bss}
\end{table}

\newpage

\begin{table}[h]
 \caption{Central values (c-value)  and bounds for unparticle couplings with $b$ quark in
the loop for $\Lambda_\U =$ 1 TeV. In the table ``-'' indicates that the central values are larger than
10 implying weak constraints which we do not list. The corresponding values for $\alpha_\Lambda$ are listed
below the constraints on couplings. In the table ``$\sim 1$.'' indicates a value very close to one.}
\vspace{0.5cm}{\scriptsize
$\begin{array}{|c|ccccc|}
\hline
d_\U & 1.1 & 1.3 & 1.5 & 1.7 & 1.9 \\
\hline

\text{$\lambda^R_{s b}\lambda^R_{b b}$ c-value}&-0.00026(0.00020)&-0.0053(0.0043)&-0.079(0.067)&-0.72(0.
63)&-1.8(1.6)\\
\text{$\alpha_\Lambda$}&0.76(0.81)&0.77(0.81)&0.78(0.8)&0.78(0.8)&0.79(0.79)\\
\text{$\lambda^R_{s b}\lambda^R_{b b}$ bound}&-0.00035(0.00030)&-0.0072(0.0063)&-0.11(0.096)&-0.99(0.90)&
-2.4(2.2)\\
\text{$\alpha_\Lambda$}&0.59(0.65)&0.6(0.65)&0.61(0.64)&0.62(0.63)&0.62(0.63)\\
\hline

\text{$\lambda^L_{sb}\lambda^L_{bb}$ c-value}&-0.0014(3.8\x{-5})&-0.029(0.00080)&-0.44(0.012)&-4.0(0.
11)&-10(0.28)\\
\text{$\alpha_\Lambda$}&\sim 1.(1.)&\sim 1.(1.)&\sim 1.(1.)&\sim 1.(1.)&\sim 1.(1.)\\
\text{$\lambda^L_{sb}\lambda^L_{bb}$ bound}&-0.0014(7.3\x{-5})&-0.030(0.0015)&-0.45(0.023)&-4.2(0.21)&
-10(0.53)\\
\text{$\alpha_\Lambda$}&\sim 1.(1.)&\sim 1.(1.)&\sim 1.(1.)&\sim 1.(1.)&\sim 1.(1.)\\
\hline

\text{$\lambda^R_{s b}\lambda^L_{b b}$ c-value}&-8.9\x{-5}(9.8\x{-5})&-0.0023(0.0025)&-0.046(0.05)&-0.
7(0.76)&-5.2(5.7)\\
\text{$\alpha_\Lambda$}&0.79(0.79)&0.79(0.79)&0.78(0.8)&0.78(0.8)&0.78(0.8)\\
\text{$\lambda^R_{s b}\lambda^L_{b b}$ bound}&-0.00013(0.00014)&-0.0032(0.0034)&-0.065(0.07)&-0.99(1.1)
&-7.4(7.9)\\
\text{$\alpha_\Lambda$}&0.62(0.63)&0.62(0.63)&0.62(0.63)&0.62(0.63)&0.62(0.63)
\\
\hline

\text{$\lambda^L_{s b}\lambda^R_{b b}$ c-value}&-1.6\x{-5}(0.00056)&-0.00039(0.014)&-0.008(0.29)&-0.12(
4.4)&-0.91(33)\\
\text{$\alpha_\Lambda$}&\sim 1.(1.)&\sim 1.(1.)&\sim 1.(1.)&\sim 1.(1.)&\sim 1.(1.)\\
\text{$\lambda^L_{s b}\lambda^R_{b b}$ bound}&-3.0\x{-5}(0.00058)&-0.00075(0.015)&-0.015(0.30)&-0.23(4.
5)&-1.7(34)\\
\text{$\alpha_\Lambda$}&\sim 1.(1.)&\sim 1.(1.)&\sim 1.(1.)&\sim 1.(1.)&\sim 1.(1.)\\
\hline

\text{$\lambda^{\YD}_{sb}\lambda^{\YD}_{bb}$ c-value}&-0.00075(0.027)&-0.011(0.42)&-0.14(5.0)&-1.1(39)&-2.4(
86)\\
\text{$\alpha_\Lambda$}&\sim 1.(1.)&\sim 1.(1.)&\sim 1.(1.)&\sim 1.(1.)&\sim 1.(1.)\\
\text{$\lambda^{\YD}_{sb}\lambda^{\YD}_{bb}$ bound}&-0.0014(0.028)&-0.022(0.43)&-0.27(5.1)&-2.1(40)&-4.5(88
)\\
\text{$\alpha_\Lambda$}&\sim 1.(1.)&\sim 1.(1.)&\sim 1.(1.)&\sim 1.(1.)&\sim1.(1.)\\
\hline

\text{$\lambda^{\YD}_{bs}\lambda^{\YD}_{bb}$ c-value}&-0.0042(0.0048)&-0.065(0.074)&-0.78(0.89)&-6.1(6.9)&-\\
\text{$\alpha_\Lambda$}&0.99(\sim 1.)&0.99(\sim 1.)&\sim 1.(1.)&\sim 1.(1.)&-\\
\text{$\lambda^{\YD}_{bs}\lambda^{\YD}_{bb}$ bound}&-0.0060(0.0066)&-0.092(0.10)&-1.1(1.2)&-8.7(9.5)&-\\
\text{$\alpha_\Lambda$}&0.98(0.99)&0.99(\sim 1.)&0.99(\sim 1.)&\sim 1.(1.)&-\\
\hline
\end{array}$}
\label{bbs}
\end{table}


\newpage

\begin{references}
\bibitem{Georgi:2007ek}
  H.~Georgi,
  Phys.\ Rev.\ Lett.\  {\bf 98}, 221601 (2007).
  H.~Georgi,
  Phys.\ Lett.\  B {\bf 650}, 275 (2007)
  [arXiv:0704.2457 [hep-ph]].



\bibitem{Chen:2007qr}
  S.~L.~Chen and X.~G.~He,
  Phys.\ Rev.\  D {\bf 76}, 091702 (2007)
  [arXiv:0705.3946 [hep-ph]];


 \bibitem{quark}
  M.~Luo and G.~Zhu,
  Phys.\ Lett.\  B {\bf 659}, 341 (2008)
  [arXiv:0704.3532 [hep-ph]];  C.~H.~Chen and C.~Q.~Geng,
  Phys.\ Rev.\  D {\bf 76}, 115003 (2007)
  [arXiv:0705.0689 [hep-ph]];  X.~Q.~Li and Z.~T.~Wei,
  Phys.\ Lett.\  B {\bf 651}, 380 (2007)
  [arXiv:0705.1821 [hep-ph]];  C.~D.~Lu, W.~Wang and Y.~M.~Wang,
  Phys.\ Rev.\  D {\bf 76}, 077701 (2007)
  [arXiv:0705.2909 [hep-ph]];  T.~M.~Aliev, A.~S.~Cornell and N.~Gaur,
  JHEP {\bf 0707}, 072 (2007)
  [arXiv:0705.4542 [hep-ph]];  C.~H.~Chen and C.~Q.~Geng,
  Phys.\ Rev.\  D {\bf 76}, 036007 (2007)
  [arXiv:0706.0850 [hep-ph]];
  R.~Zwicky,
  Phys.\ Rev.\  D {\bf 77}, 036004 (2008)
  [arXiv:0707.0677 [hep-ph]];
  R.~Mohanta and A.~K.~Giri,
  Phys.\ Rev.\  D {\bf 76}, 075015 (2007)
  [arXiv:0707.1234 [hep-ph]];
  C.~S.~Huang and X.~H.~Wu,
  Phys.\ Rev.\  D {\bf 77}, 075014 (2008)
  [arXiv:0707.1268 [hep-ph]];  A.~Lenz,
  Phys.\ Rev.\  D {\bf 76}, 065006 (2007)
  [arXiv:0707.1535 [hep-ph]];
  R.~Mohanta and A.~K.~Giri,
  Phys.\ Rev.\  D {\bf 76}, 057701 (2007)
  [arXiv:0707.3308 [hep-ph]];
  C.~H.~Chen and C.~Q.~Geng,
  Phys.\ Lett.\  B {\bf 661}, 118 (2008)
  [arXiv:0709.0235 [hep-ph]];
  T.~M.~Aliev and M.~Savci,
  Phys.\ Lett.\  B {\bf 662}, 165 (2008)
  [arXiv:0710.1505 [hep-ph]];  S.~L.~Chen, X.~G.~He, X.~Q.~Li, H.~C.~Tsai and Z.~T.~Wei,
  arXiv:0710.3663 [hep-ph];
  R.~Mohanta and A.~K.~Giri,
  Phys.\ Lett.\  B {\bf 660}, 376 (2008)
  [arXiv:0711.3516 [hep-ph]];  Y.~f.~Wu and D.~X.~Zhang,
  arXiv:0712.3923 [hep-ph];  C.~H.~Chen, C.~S.~Kim and Y.~W.~Yoon,
  arXiv:0801.0895 [hep-ph];
  V.~Bashiry,
  arXiv:0801.1490 [hep-ph];
  M.~J.~Aslam and C.~D.~Lu,
  arXiv:0802.0739 [hep-ph];  Y.~Wu and D.~X.~Zhang,
  arXiv:0804.1843 [hep-ph].

\bibitem{lepton}

  T.~M.~Aliev, A.~S.~Cornell and N.~Gaur,
  Phys.\ Lett.\  B {\bf 657}, 77 (2007)
  [arXiv:0705.1326 [hep-ph]];
  D.~Choudhury, D.~K.~Ghosh and Mamta,
  Phys.\ Lett.\  B {\bf 658}, 148 (2008)
  [arXiv:0705.3637 [hep-ph]];  E.~O.~Iltan,
  arXiv:0711.2744 [hep-ph];  E.~O.~Iltan,
  arXiv:0801.0301 [hep-ph];
  A.~Hektor, Y.~Kajiyama and K.~Kannike,
  arXiv:0802.4015 [hep-ph].

\bibitem{Ding:2008zza}
  G.~J.~Ding and M.~L.~Yan,
  Phys.\ Rev.\  D {\bf 77}, 014005 (2008).
  G.~j.~Ding and M.~L.~Yan,
  arXiv:0709.3435 [hep-ph].


\bibitem{new}
  J.~L.~Hewett,
  arXiv:hep-ph/9406302;
  X.~G.~He, T.~D.~Nguyen and R.~R.~Volkas,
  Phys.\ Rev.\  D {\bf 38}, 814 (1988);
  C.~K.~Chua, X.~G.~He and W.~S.~Hou,
  Phys.\ Rev.\  D {\bf 60}, 014003 (1999)
  [arXiv:hep-ph/9808431];
  X.~G.~He and B.~McKellar,
  Phys.\ Lett.\  B {\bf 320}, 165 (1994)
  [arXiv:hep-ph/9309228].

\bibitem{HFAG:2008}
   The Heavy Flavor Averaging Group (HFAG), 2008. http://www.slac.stanford.edu/xorg/hfag/ rare/winter08/radll/btosg.pdf.
\bibitem{Misiak:2007}
  M.~Misiak \emph{et al.}, Phys.\ Rev.\ Lett.\  {\bf 98}, 022002 (2007).

\bibitem{Deshpande:2007jy}
  N.~G.~Deshpande, X.~G.~He and J.~Jiang,
  Phys.\ Lett.\  B {\bf 656}, 91 (2007)
  [arXiv:0707.2959 [hep-ph]].

\bibitem{Fox:2007sy}
  P.~J.~Fox, A.~Rajaraman and Y.~Shirman,
  Phys.\ Rev.\  D {\bf 76}, 075004 (2007)
  [arXiv:0705.3092 [hep-ph]].


\bibitem{pdg} W.-M. Yao et al., (Particle Data Group), J. Phys. {\bf G33}, 1(2006).


\bibitem{Buras:1993xp}
  A.~J.~Buras, M.~Misiak, M.~Munz and S.~Pokorski,
  Nucl.\ Phys.\  B {\bf 424}, 374 (1994)
  [arXiv:hep-ph/9311345].

\bibitem{polarization} T.~Mannel and S.~Recksiegel,
  J.\ Phys.\ G {\bf 24}, 979 (1998)
  [arXiv:hep-ph/9701399];
C.~K.~Chua, X.~G.~He and W.~S.~Hou,
  Phys.\ Rev.\  D {\bf 60}, 014003 (1999)
  [arXiv:hep-ph/9808431].

\end{references}
\end{document}